\documentclass[preprint,aps ,nofootinbib]{revtex4}
\usepackage{graphicx}
\pdfoutput=1
\usepackage{epsfig}
\usepackage{amsmath}
\usepackage{amsfonts}
\usepackage{amssymb}
\usepackage{color}
\usepackage{dcolumn}
\usepackage{hyperref}
\usepackage{multirow}
\usepackage{booktabs}
\providecommand{\U}[1]{\protect\rule{.1in}{.1in}}

\newcommand{\f}{\begin{equation}}
\newcommand{\ff}{\end{equation}}
\newcommand{\fa}{\begin{eqnarray}}
\newcommand{\ffa}{\end{eqnarray}}

\begin{document}
\title{Regular black holes with sub-Planckian curvature }
\author{Yi Ling $^{1,2}$}
\email{lingy@ihep.ac.cn}
\author{Meng-He Wu$^{3,4}$}
\email{mhwu@sues.edu.cn} \affiliation{$^1$Institute of High Energy
Physics, Chinese Academy of Sciences, Beijing 100049, China\\ $^2$
School of Physics, University of Chinese Academy of Sciences,
Beijing 100049, China \\
$^3$ School of Mathematics, Physics and Statistics, Shanghai University of Engineering Science, Shanghai 201620, China \\
$^4$Center of Application and Research of Computational Physics, Shanghai University of Engineering Science, Shanghai 201620, China}

\begin{abstract}
We construct a sort of regular black holes  with a sub-Planckian
Kretschmann scalar curvature. The metric of this sort of regular
black holes is characterized by an exponentially suppressing
gravity potential as well as an asymptotically Minkowski core. In
particular, with different choices of the potential form, they can
reproduce the metric of Bardeen/Hayward/Frolov black hole at large
scales. The heuristical derivation of this sort of black holes is
performed based on the generalized uncertainty principle over
curved spacetime which includes the effects of tidal force on any
object with finite size which is bounded below by the minimal
length.

\end{abstract}
\maketitle
\section{Introduction}
It is widely believed that quantum gravitational effects would
remove the curvature singularity of a black hole. Before a
complete theory of quantum gravity could be established, people
have extensively investigated various non-singular black holes at
the phenomenological level, which usually are not solutions to the
vacuum Einstein equations. Instead some exotic matter fields must
be introduced which in general violate some energy conditions in
general relativity.

Regular black holes were originally proposed to avoid the
singularity of ordinary black holes\cite{Bardeen:1968,Hayward:2005gi,Frolov:2014jva,Ansoldi:2008jw,Bonanno:2000ep,KalyanaRama:2001xd,Benczik:2002tt,Nicolini:2005vd,Myung:2007qt,Balart:2014jia,Amir:2016cen}, which includes the well known Bardeen black hole, Hayward black hole as well as Frolov black hole\cite{Bardeen:1968,Hayward:2005gi,Frolov:2014jva,Ansoldi:2008jw}.
The Bardeen black hole is the first regular black hole that obeys the weak energy condition. It was originally proposed as a counterexample to prove the possibility of singularities in black hole space without the need to assume global Cauchy hypersurfaces or strong energy conditions\cite{Borde:1996df}. Later, in order to describe the formation and evaporation of black holes, the Hayward  black holes were constructed\cite{Hayward:2005gi,Pedraza:2020uuy}.  Ref.\cite{Neves:2014aba,Maluf:2018lyu} constructed a general class of black holes and included Bardeen and Hayward black hole.
The most important property of these
regular black holes is that its Kretschmann scalar curvature is
finite everywhere. In particular, for Bardeen/Hayward/Frolov black holes,  the singularity which
appears at $r=0$ in ordinary black holes now is replaced by the de-Sitter spacetime. That is to say, the asymptotic metric of those
regular black holes near the center ($r\rightarrow 0$) becomes
\begin{equation}\label{GFGP}
F(r)=1-\frac{r^2}{l^2},
\end{equation}
which is nothing but the solution to the Einstein
equation with the cosmological constant $\Lambda=3/l^2$, where $l$ is a constant related to the Planck length $l_p$ and $\Lambda$ may be understood as the effective
cosmological constant at small distance\cite{Hayward:2005gi}.
Therefore, they are
also called as regular black holes with de-Sitter core.
In Ref.\cite{Xiang:2013sza}, the authors analyzed the general conditions
that would lead to a finite Kretschmann scalar and originally
proposed a black hole solution with an asymptotically Minkowski
core. In contrast to all the previous regular black holes as
mentioned above, as $r\rightarrow 0$, the function $F(r)$ in the
metric becomes $ F(r)=1$ such that the singularity of the black
hole is replaced by Minkowski spacetime rather than de-Sitter
spacetime. Thus we call this solution as the regular black hole
with Minkowski core.
 This kind of regular black hole is featured by an
exponentially suppressing potential and a vanishing Hawking
temperature at the final stage of evaporation, in contrast to the
phenomenon that the Hawking temperature becomes divergent for
classical black holes, as well to the phenomenon that the Hawking
temperature takes a maximal value staying at the Planck
energy level for semi-classical black holes where the quantum
effects of gravity such as the generalized uncertainty principle
(GUP)\cite{Maggiore:1993rv,Garay:1995,Scardigli:1999jh,Adler:2001vs}
or modified dispersion relations (MDR) are taken into
account\cite{Amelino-Camelia:2004uiy,Amelino-Camelia:2005zpp,Ling:2005bq,Han:2008sy}.
Thus, this regular black hole solution provides more realistic
picture for the final stage of black hole evaporation and more
reasonable behavior of the black hole remnant. Subsequently this
solution has been generalized into a kind of regular black holes
with different forms of the exponential potential in
\cite{Culetu:2013fsa,Culetu:2014lca,Rodrigues:2015ayd,Simpson:2019mud,Ghosh:2014pba,Ghosh:2018bxg}.
Nevertheless,  we notice that  for the regular black holes proposed in \cite{Xiang:2013sza}, the  maximal value of Kretschmann scalar curvature depends on the mass of
black hole, which is in contrast to  Bardeen/Hayward/Frolov black holes where the maximal value of Kretschmann scalar curvature is mass independent. Explicitly, once the parameter $\alpha$ is given in \cite{Xiang:2013sza}, which is supposed to be fixed by the quantum
effects of gravity, the maximal value is proportional to the
square of the mass. It means that Kretschmann scalar curvature is
not bounded above, but can exceed the Planck mass density
$(M_p^4)$ easily by increasing the mass of the black hole, which
of course is not reasonable or expectable from quantum gravity
point of view. Or in another word, it implies that the metric for
 previous regular black holes with Minkowski core  makes sense
only for black holes with small mass at the Planck scale, or for
the final stage of black hole evaporation.
 Additionally, the regular black holes with Minkowski cores also have interesting feature. The stress-energy tensor vanishes at the centre. This implies that the physics of this region is greatly simplified compared to the black holes with de-Sitter core\cite{Simpson:2019mud,Berry:2020ntz}.

The purpose of this paper is twofold. Firstly, we intend to
construct a new sort of regular black holes with Minkowski core,
getting rid of the shortcoming that the Kretschmann scalar
curvature is not bounded by the Planck mass density, such that the
metric is applicable to the black hole with arbitrarily large
mass, and thus applicable to all the stage of black hole
evaporation from the beginning to the end. Secondly, we intend to
disclose a closer relation between regular black holes with
Minkowski core and those with de-Sitter core. Specifically, we will
demonstrate that with different choices of the potential form,
this sort of black holes can reproduce the metric of
Bardeen/Hayward/Frolov black hole at large scales. In  this
situation, these two kinds of black holes may have distinct
behavior only at the late stage of evaporation.

This paper is organized as follows. In next section we will
present the general setup for this sort of regular black holes
with spherical symmetry, and then in section three we find the
condition that Kretschmann scalar curvature can be bounded above
and then show a specific example in comparison with the solution
proposed in \cite{Xiang:2013sza}. In section four and five, we
will demonstrate that for specifical choice of the potential from,
the regular black holes have the similar behavior as
Bardeen/Hayward black holes at large scales, respectively.
Finally, we will argue how the consideration of generalized
uncertainty principle over curved spacetime could lead to this
sort of black holes in a heuristical manner. We point out that the
effect of tidal force on any object with finite size bounded by
the minimal length plays an essential role in this modification.

\section{The general setup for static regular black holes with spherical symmetry}

 Firstly, to construct the regular black hole at the phenomenological level, the key point is to obtain a finite value for Kretschmann scalar curvature such that the singularity is avoided. For this purpose we consider a static spherically symmetric black hole with a general form of the metric
\begin{equation}
d s^2=-\left(1+2 \phi_1\right) d t^2+\left(1+2 \phi_2\right)^{-1} d r^2+r^2 d \Omega,
\end{equation}
where $\phi_1=\phi_1(r), \phi_2=\phi_2(r)$ are gravitational potentials. As shown in \cite{Xiang:2013sza}, it was found that to get rid of the singularity, one essential condition  is that as $r \rightarrow 0$, the  asymptotic behaviors of modified gravitational potential must take the form
\begin{equation}
\begin{aligned}
& \phi_1 \rightarrow r^l, l \geq 2, \\
& \phi_2 \rightarrow r^s, s \geq 2.
\end{aligned}
\end{equation}
For simplicity, in this paper we consider the case $\phi_1=\phi_2$ such that the  metric takes the following form
\begin{equation}\label{Eq_met}
d s^{2}=- F(r) d t^{2}+\frac{1}{F(r)} d r^{2}+r^{2} d
\Omega^{2},
\end{equation}
with
\begin{equation}
 F(r)=1+2 \psi(r).
\end{equation}
Typically, if $\psi(r)=-GM/r$, then it is nothing but the
Schwarzschild black hole.
Now, in order to include the strong quantum effects
of gravity which would modify the singularity behavior at the
Planck scale, we assume $F(r)$ would be a general form, with the
requirement that it goes back to Schwarzschild black hole at
asymptotical infinity. The location of the horizon $r_h$ is
determined by $g^{rr}=F(r_h)=0$.

From Einstein field equations $G_{\nu}^{\mu}=8 \pi T_{\nu}^{\mu}$,
one can derive the effective stress-energy tensor which is given
by $T_{\nu}^{\mu}=\frac{1}{4\pi r} diag \{\frac{\psi}{r}+\psi',
\frac{\psi}{r}+\psi',
\psi'+\frac{r}{2}\psi'',\psi'+\frac{r}{2}\psi'' \}$.

Furthermore, it is straightforward to derive Kretschmann scalar
curvature which is given by
\begin{equation}
\begin{aligned}
K &=R^{\mu \nu \rho \lambda} R_{\mu \nu \rho \lambda} \\
&=\frac{16 \psi (r)^2}{r^4}+\frac{16 \psi '(r)^2}{r^2}+4 \psi ''(r)^2.
\end{aligned}
\end{equation}

The Hawking temperature of the  black hole and the luminosity are
respectively given by
\begin{equation}
\begin{aligned}
 T&=\frac{F'(r_h)}{4 \pi }=\frac{\psi '(r_h)}{2 \pi },\\
L &= \frac{  \sigma r_{h} ^{2}  F'(r_h)^4}{64 \pi ^3}=\frac{\sigma
r_{h} ^{2}  \psi '(r)^4}{4 \pi ^3}.
\end{aligned}
\end{equation}

Now we propose a new sort of regular black holes with asymptotically
Minkowski core which is constructed with an exponentially
suppressing form of the gravitational potential, where $\psi(r)$
is specified as
\begin{equation}\label{GFGP}
\psi=-\frac{GM}{r} e^{-\alpha_0 (GM)^x l_{p}^{n-x} / r^{n}},
\end{equation}
with $\alpha_0>0$, $n> x\geq 0$ and $n\geq 1$. In addition, all the parameters $\alpha_0$, $n$ and $x$ are understood as dimensionless. Throughout this paper, for simplicity we ignore the factor difference of $G$ and $l_p$ by setting $G=l_p^2=1$.

In comparison with  previous regular black holes with Minkowski
core in \cite{Xiang:2013sza},    the
exponential form in our formalism may depend on the mass of black hole, which plays
an essential role in suppressing the maximal value of Kretschmann
scalar curvature to be sub-Planckian.   Moreover, for specific values of $x$ and $n$, one can establish a one-to-one correspondence between  the regular black hole with Minkowski core and that with de-Sitter core at large scale.   We will explicitly
demonstrate this feature in next sections. Obviously, if $x=0$,
then it goes back to the regular black hole proposed by Li {\it
et.al.}\cite{Xiang:2013sza}.

\section{The regular black hole with sub-Planckian Kretschmann scalar curvature}
In this section we will construct a regular black hole with
sub-Planckian Kretschmann scalar curvature by setting $x=1$ and
$n=2$.

\begin{figure} [t]
  \center{
  \includegraphics[scale=0.35]{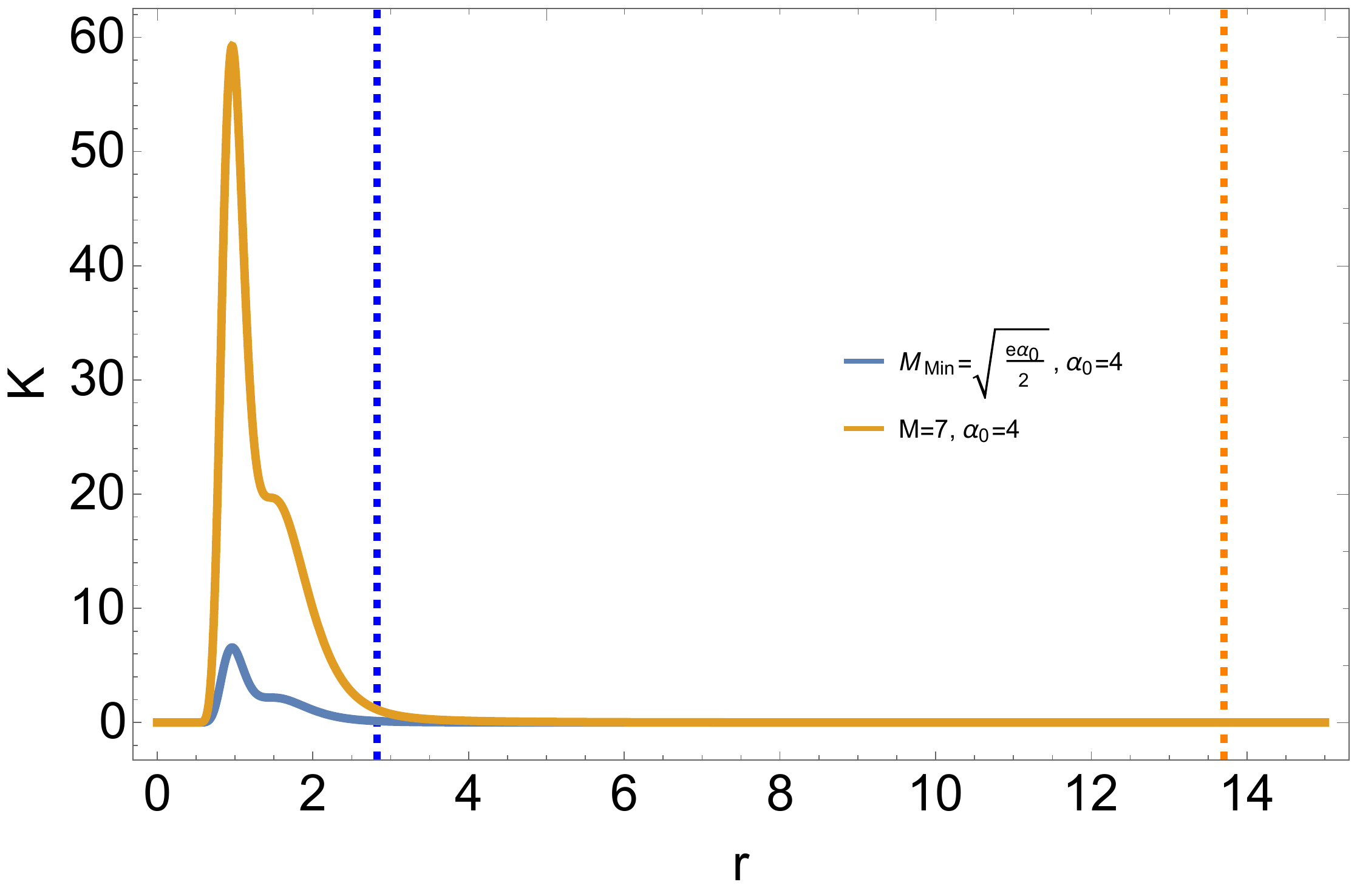}\ \hspace{0.05cm}
  \includegraphics[scale=0.36]{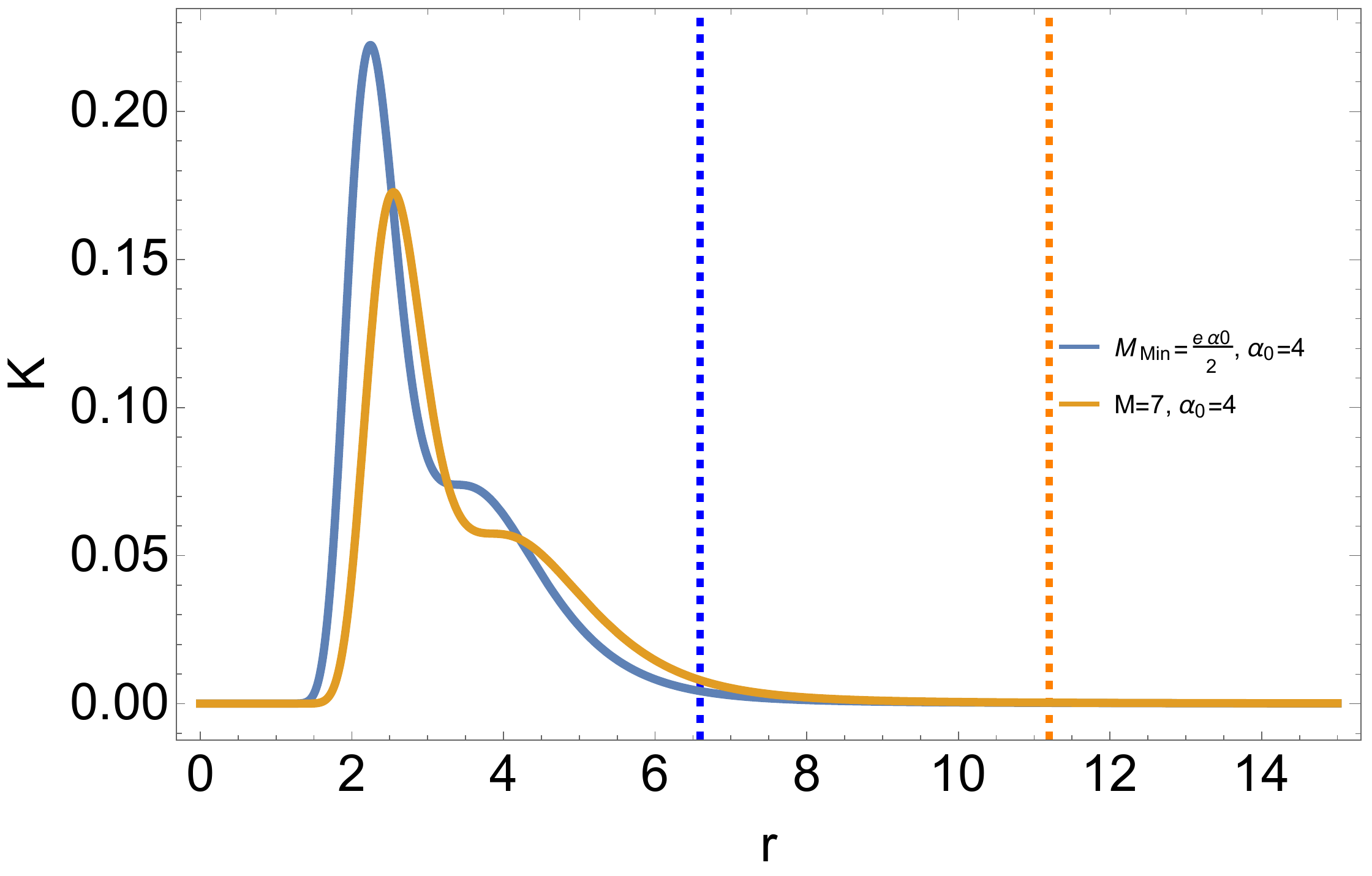}\ \hspace{0.05cm}
  \caption{\label{fig1}  The Kretschmann scalar curvature $K$  as the functions of the radial coordinate $r$   for $x=0$(left) and $x=1$(right).}}
\end{figure}

First of all, it is helpful to understand why all the previous
regular black holes with $x=0$ lead to a huge Kretschmann scalar
curvature whenever the mass of black hole becomes large. For
instance, for the regular black hole proposed by Li {\it et.al.}\cite{Xiang:2013sza},
the gravity potential is  $\psi=-\frac{M}{r} e^{-\alpha_0
/r^{2}}$, and Kretschmann scalar curvature is given by
\begin{equation}\label{KSC1}
\begin{aligned}
K &=\frac{16 M^2 e^{-\frac{2 \alpha_0 }{r^2}}}{r^6}(3-\frac{14
\alpha_0
   }{r^2}+\frac{33 \alpha_0
^2}{r^4}-\frac{20 \alpha_0 ^3}{r^6}+\frac{4 \alpha_0 ^4}{r^8}),
\end{aligned}
\end{equation}
which is finite everywhere. In particular, near the center
$r\rightarrow 0$, $K\rightarrow 0$, which is in contrast to the
standard Schwarzschild black hole in which $K$ becomes divergent.
However, one notices that once $\alpha_0$ is fixed as it should be
from the quantum gravity point of view, the maximum value of $K$
is proportional to $M^2$, which means this value may easily exceed
the Planck mass density for large black holes, namely $K>1$ with
the unit of $m_p^4$. We show this behavior in the right plot of
Fig.\ref{fig1}. Of course this feature is not satisfactory if one
insists that all the reasonable quantities should be sub-Planckian
due to the quantum gravity effects.

On the other hand, if we assume that the exponential factor could
be mass dependent, then such situation would change dramatically.
As a matter of fact, if we set $n=2$ and replace $\alpha_0$ by
$\alpha_0 M^x$, then one obtains $K$ with the similar expression
as Eq.(\ref{KSC1}), simply replacing $\alpha_0$ by $\alpha_0 M^x$.
A simple algebra shows that now the maximal value of $K$ is
proportional to $M^{2-3x}/\alpha_0 ^3$. Therefore, if we demand
$x\geq 2/3$, then the maximal value would inversely be
proportional to the mass. It is this observation that leads us to
propose such generalized model. Without loss of generality, we
will set $x=1$ in this section and compare this regular black
holes with the previous one \cite{Xiang:2013sza}, which is
obtained by setting $x=0$.

Firstly, we present the basic properties of the regular black hole when we take $x=1$. From $F(r_h)=0$, one obtains the relation between the location of horizon $r_h$ and the mass $M$ as
\begin{equation}
2M=r_h e^{\alpha_0 M / r_{h}^{2}} .
\end{equation}
Then  the radius of the horizon $r_h$  can be expressed as a function of mass
\begin{equation}
r_h=2 M \sqrt{\frac{\theta}{W(\theta)}}, \quad
\theta=-\frac{\alpha_0 }{2 M},
\end{equation}
where $W(\theta)$ is the Lambert-W function. As discussed in \cite{Simpson:2019mud,Zeng:2022tzh}, the real-valued $W(\theta)$ with negative arguments has two branches, corresponding to the outer and inner horizons of the black hole, respectively.
The inner horizon of the black hole is located at $r=2 M \sqrt{\frac{\theta}{W_{-1}(\theta)}}$ while the outer horizon is located at  $r=2 M \sqrt{\frac{\theta}{W_{0}(\theta)}}$.
In this paper, we are concerned with the thermodynamical properties of the black hole which are closely related to the outer horizon,  thus we will concentrate on the analysis of the outer horizon, where the Lambert-W function is given by:
\begin{equation}
W_{0}(\theta)=\sum_{n=1}^{\infty} \frac{(-n)^{n-1}}{n !} \theta^{n},
\end{equation}
with $W_{0}(\theta)\geq-1$\cite{Valluri:2000zz}.

The above equation is quite similar to
the one for $x=0$, but now the variable $\theta =-\frac{\alpha_0
}{2 M}$, rather than $\theta =-\frac{\alpha_0 }{2 M^2}$ in
\cite{Xiang:2013sza}. Moreover, as pointed out in
\cite{Xiang:2013sza}, a real W requires $\theta \geq -e^{-1}$
such
that we have the lowest bound for the mass of the black hole
\begin{equation}
M\geq \frac{e\alpha_0   }{2}.
\end{equation}
When this bound is saturated, the black hole is characterized by
the minimal radius of the horizon $r_h=\sqrt{e}\alpha_0$, which
may be treated as the remnant of the black hole evaporation, as we
will elaborate it as below. We remark that when $M = \frac{e\alpha_0   }{2}$, the inner and outer horizons of the black hole are merged and the black hole becomes extremal since the temperature goes to zero. Such a remnant may be viewed as a candidate for the dark matter\cite{Adler:2001vs}.

Now we compare the distinct behavior of Kretschmann scalar
curvature for these two regular black holes, namely $x=0$ and
$x=1$. We plot $K(r)$ as the function of the radius for different
masses $M$ in Fig.\ref{fig1}. As we expect, the maximal value of
Kretschmann scalar curvature $K_{max}$ increases dramatically with
the increase of the mass for $x=0$,  while for $x=1$, it decreases
with $M$ indeed. We further find that the relation $K_{max}
\propto M^2$ for $x=0$, and $K_{max} \propto 1/M$ for $x=1$ can be
justified by numerical analysis. Therefore, for $x=1$ we may fix
the parameter $\alpha_0$ to have $K_{max}<1$ for $M=M_{min}$, then
it is guaranteed that Kretschmann scalar curvature is always
bounded from above for arbitrary mass $M$, as we illustrate in the
right plot of Fig.\ref{fig1}.

\begin{figure} [t]
  \center{
  \includegraphics[scale=0.41]{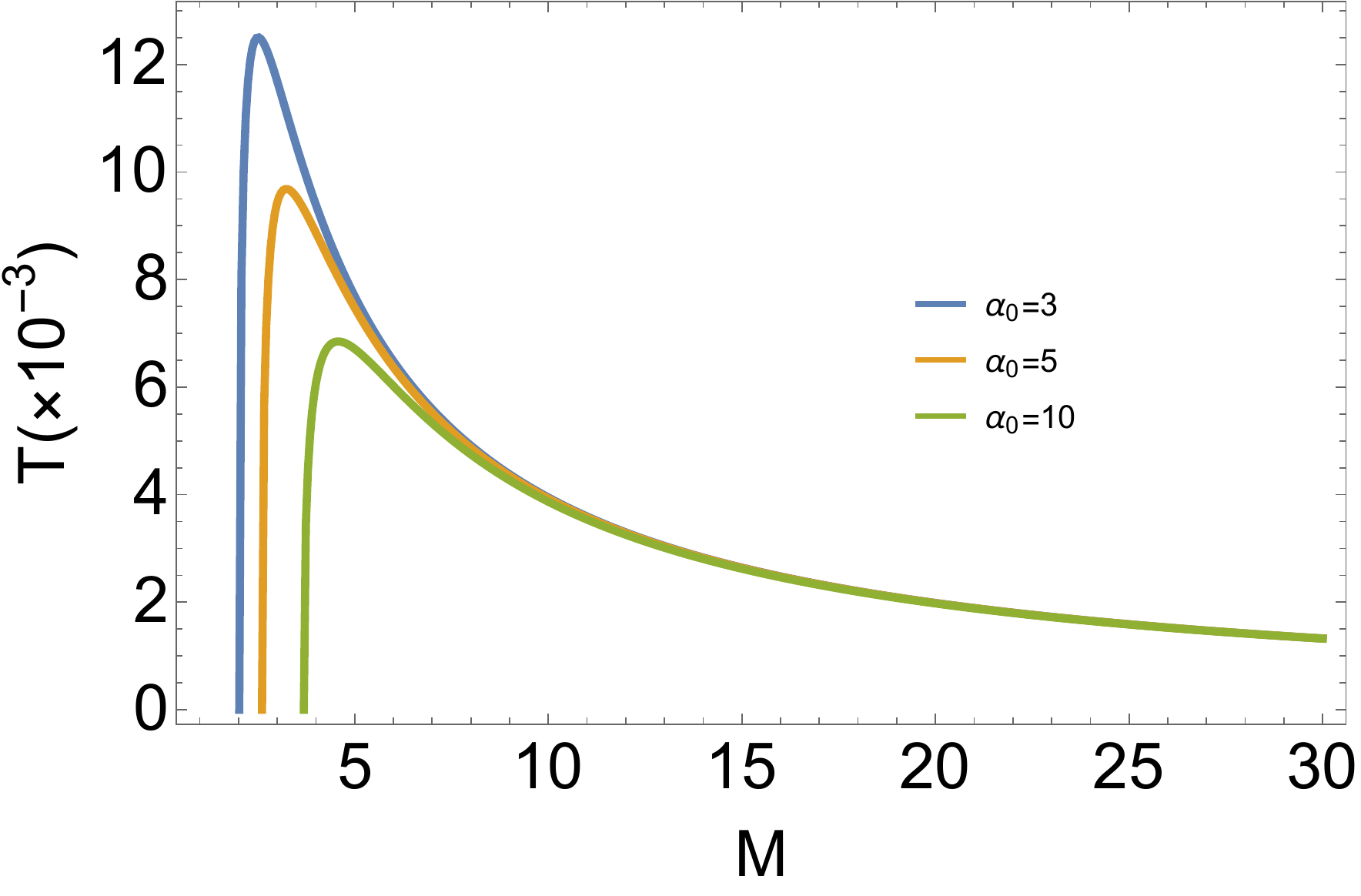}\ \hspace{0.05cm}
  \includegraphics[scale=0.4]{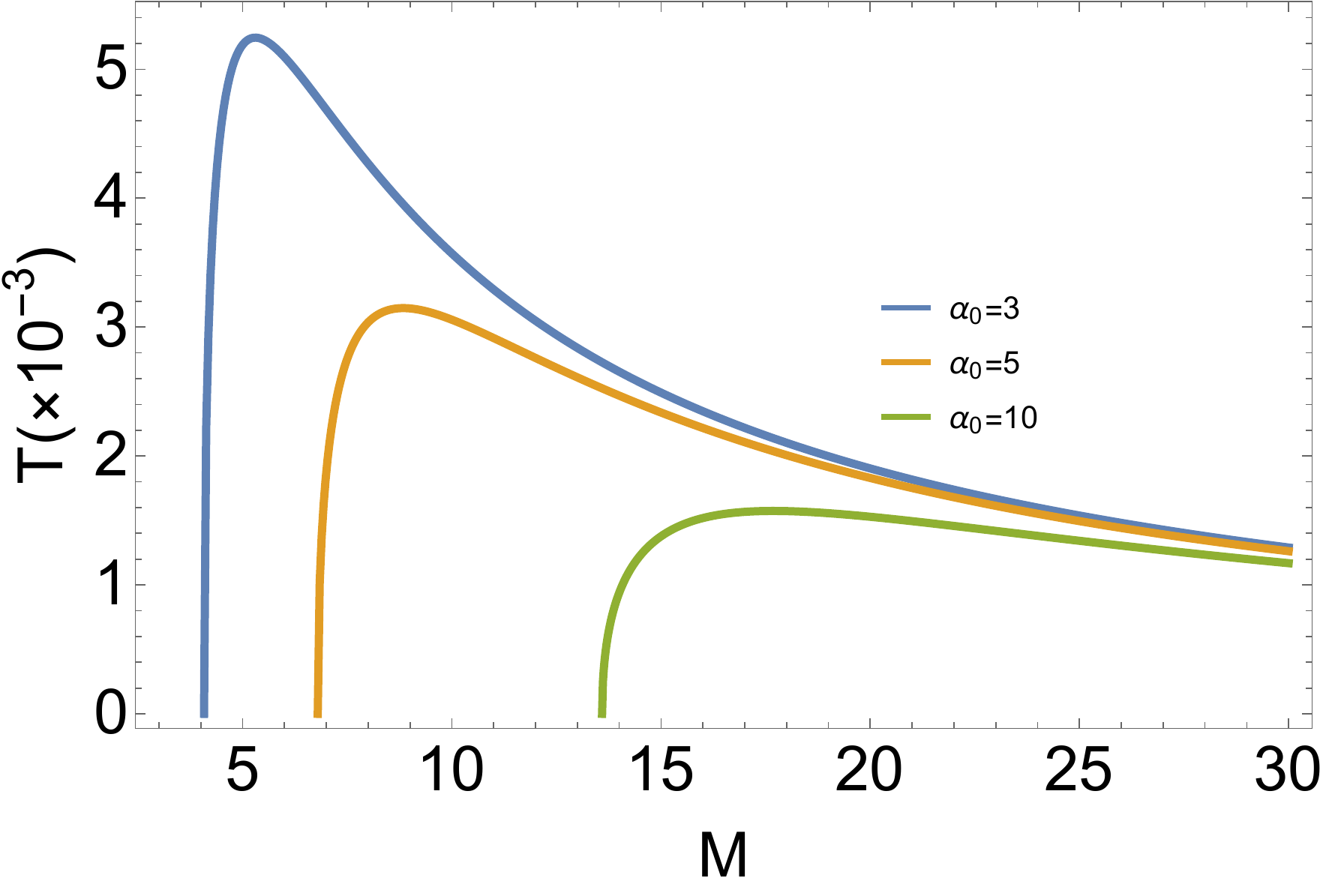}\ \hspace{0.05cm}
  \caption{\label{fig2} The Hawking temperature $T$  as the functions of  $M$ for $x=0$(left) and $x=1$(right) respectively.}}
\end{figure}

Next it is also interesting to take a look at the thermodynamical
behavior of these two regular black holes. In parallel with the
analysis presented in \cite{Xiang:2013sza}, it is straightforward
to derive the Hawking temperature $T$, the heat capacity $C\equiv
\frac{d M}{d T}$ as well as the luminosity $L$ as the function of
the mass $M$
\begin{equation}
\begin{aligned}\label{Eq_tcl}
T&=\frac{W_0+1}{8 \pi  M}  \sqrt{\frac{W_0}{\theta }},\\
C &=-\frac{16 \pi  M^2 (W_0+1)}{2+W_0(5+W_0)}\sqrt{\frac{\theta}{W_0 }}, \\
L&=\frac{\sigma   (W_0+1)^4}{256 \pi ^3   M^2}\frac{W_0}{\theta }.
\end{aligned}
\end{equation}
The temperature as well as the luminosity has the same expression
for cases $x=0$ and $x=1$, but we stress that since the variable
$\theta$ has a different relation with the mass, they exhibit
distinct behavior with the mass $M$, as illustrated in Fig.\ref{fig2}.
For $x=0$, the temperature goes to zero at the final stage
$W_0\rightarrow -1$ ($M=\sqrt{\frac{e\alpha_0}{2}}$ or
$r_h=\sqrt{2\alpha_0}$), and the temperature always reaches its
maximal value $T_{max}=1/(6\pi\sqrt{6\alpha_0})$  at $W_0=-1/3$
($M=\sqrt{\frac{3e^{1/3}}{2}\alpha_0}$ or $r_h=\sqrt{6\alpha_0}$);
while for $x=1$, the temperature goes to zero at the final stage
$W_0\rightarrow -1$ as well ($M=\frac{e\alpha_0}{2}$ or
$r_h=\sqrt{e}\alpha_0$), but reaches its maximal value at
$W_0=(-5+\sqrt{17})/2$. We also find $T_{max}\propto 1/\alpha_0$ and
$M|_{T=T_{max}}\propto \alpha_0$, implying that during the
evaporation, the temperature may reach its maximal value at larger
scale and then go down to zero in a longer time period if
$\alpha_0>1$. In addition, the remnants at zero temperature correspond to the extremal limit of the black hole.

Finally, we remark that for both black holes, the capacity is
vanishing at the final stage of evaporation, which is obvious to
see in Eq.(\ref{Eq_tcl}) as $W\rightarrow -1$. In addition,  the
capacity of both black holes undergoes a transition form $C<0$ to
$C>0$ at $T=T_{max}$.

\section{The regular black hole corresponding to Bardeen black hole at large scale}

In this section we present the regular black hole corresponding to
Bardeen black hole at large scales by appropriately choosing the
parameter $x$  with $n=2$. Recall that for Bardeen black hole, the
maximum of Kretschmann scalar curvature $K$ is independent of the
mass of black hole. Remarkably, as we analyzed in the previous
section, if we set $x=2/3$, then the maximum value of $K$ is
independent of mass $M$ as well. By virtue of this observation, we
intend to consider the regular black hole with $x=2/3$ and $n=2$.
In this case, the radius of the horizon $r_h$ is given by
\begin{equation}
r_h=2 M \sqrt{\frac{\theta}{W_0(\theta)}}, \quad
\theta=-\frac{\alpha_0 }{2 M^{4/3}}.
\end{equation}
The low bound for the mass of the black hole is
\begin{equation}
M\geq \left(\frac{e}{2}\right)^{3/4} \alpha _0^{3/4}.
\end{equation}
The Hawking temperature $T$ and the luminosity $L$ maintain the
same form as shown in Eq.(\ref{Eq_tcl}), while the heat capacity $C$
becomes
\begin{equation}
\begin{aligned}
C &=-\frac{24 \pi  M^2 (W_0+1)}{3+W_0(8+W_0)}\sqrt{\frac{\theta}{W_0 }}.
\end{aligned}
\end{equation}

Next we show that this regular black hole reproduces the Bardeen
metric at large scales. On one hand, for large $r\gg
\sqrt{\alpha_0} M^{1/3}$, the function $F(r)$ in the metric
behaves
\begin{equation}
 F(r)=1+2 \psi(r)= 1-\frac{2M}{r} e^{-\alpha_0 M^{2/3}  / r^{2}}\cong 1-\frac{2M}{r} (1-\frac{\alpha_0 M^{2/3}}{r^{2}}+...).
\end{equation}
On the other hand, for the Bardeen regular black hole, the
gravitational potential $\psi(r)$ is specified as \footnote{We remark that when the dimension is restored, the expression of $\psi(r)$ is $\psi(r)=-\frac{ G M r^2}{\left(\frac{2}{3} \alpha _0
(GM)^{2/3}l_{p}^{4/3}+r^2\right){}^{3/2}}$. }
\begin{equation}
\psi(r)=-\frac{M r^2}{\left(\frac{2}{3} \alpha _0
M^{2/3}+r^2\right){}^{3/2}}.
\end{equation}
Therefore, at large scales the function $F(r)$ behaves
\begin{equation}
 F(r)=1+2 \psi(r)= 1-\frac{2M r^2}{\left(\frac{2}{3} \alpha _0
M^{2/3}+r^2\right){}^{3/2}}\cong 1-\frac{M}{r} (1-\frac{\alpha_0
M^{2/3}}{r^{2}}+...),
\end{equation}
which is identical to the regular black hole with $x=2/3$ at large
scale, indeed.

Nevertheless, these two black holes have different cores. One has
asymptotically Minkowski core, while the other has asymptotically
de-Sitter core. Thus it is interesting to compare the behavior of
Kretschmann scalar curvature as well as thermodynamical properties
for these two regular black holes. For compactness, we intend to
summarize their key differences as the following list

\begin{figure} [t]
  \center{
  \includegraphics[scale=0.315]{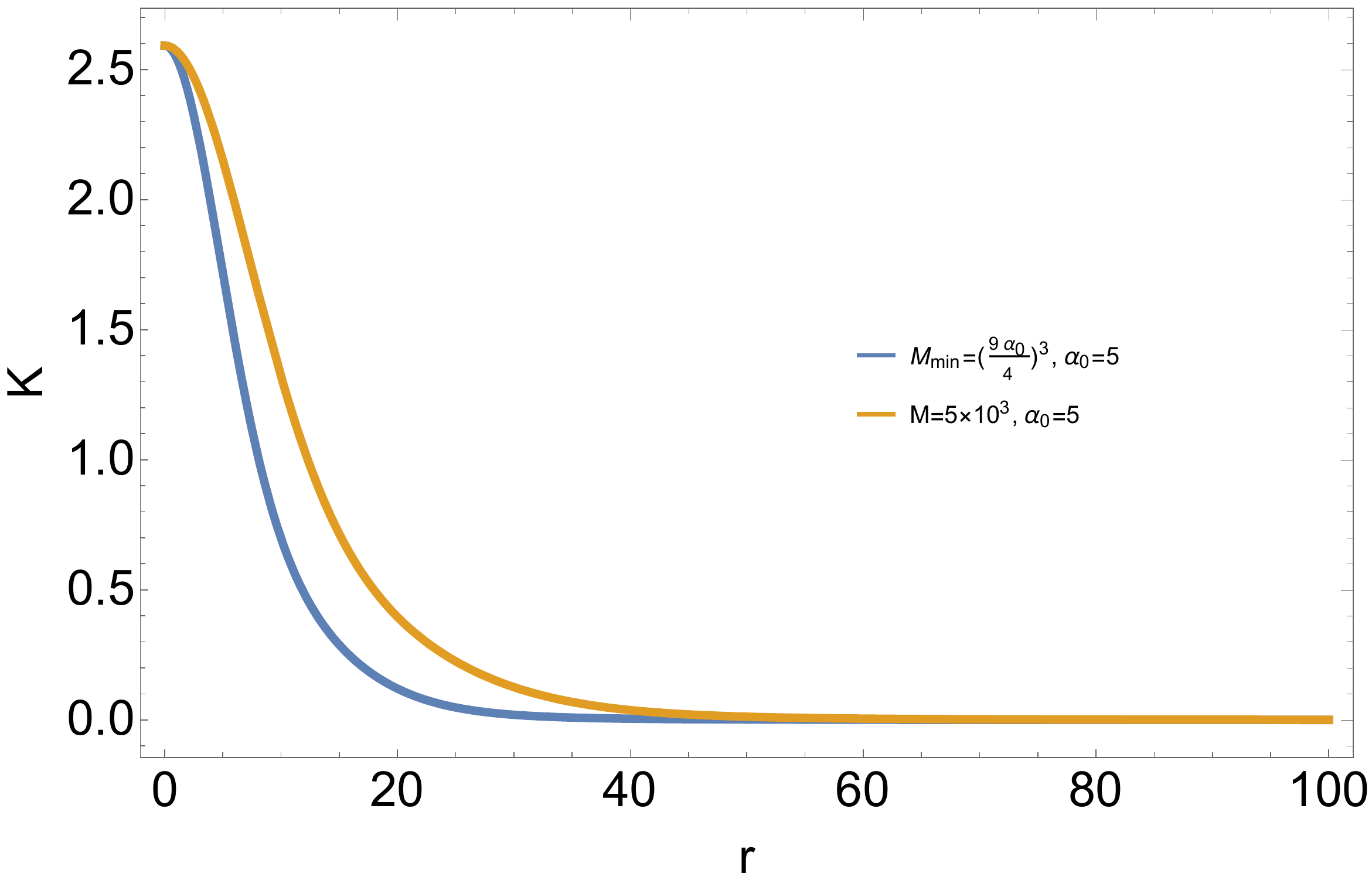}\ \hspace{0.05cm}
  \includegraphics[scale=0.31]{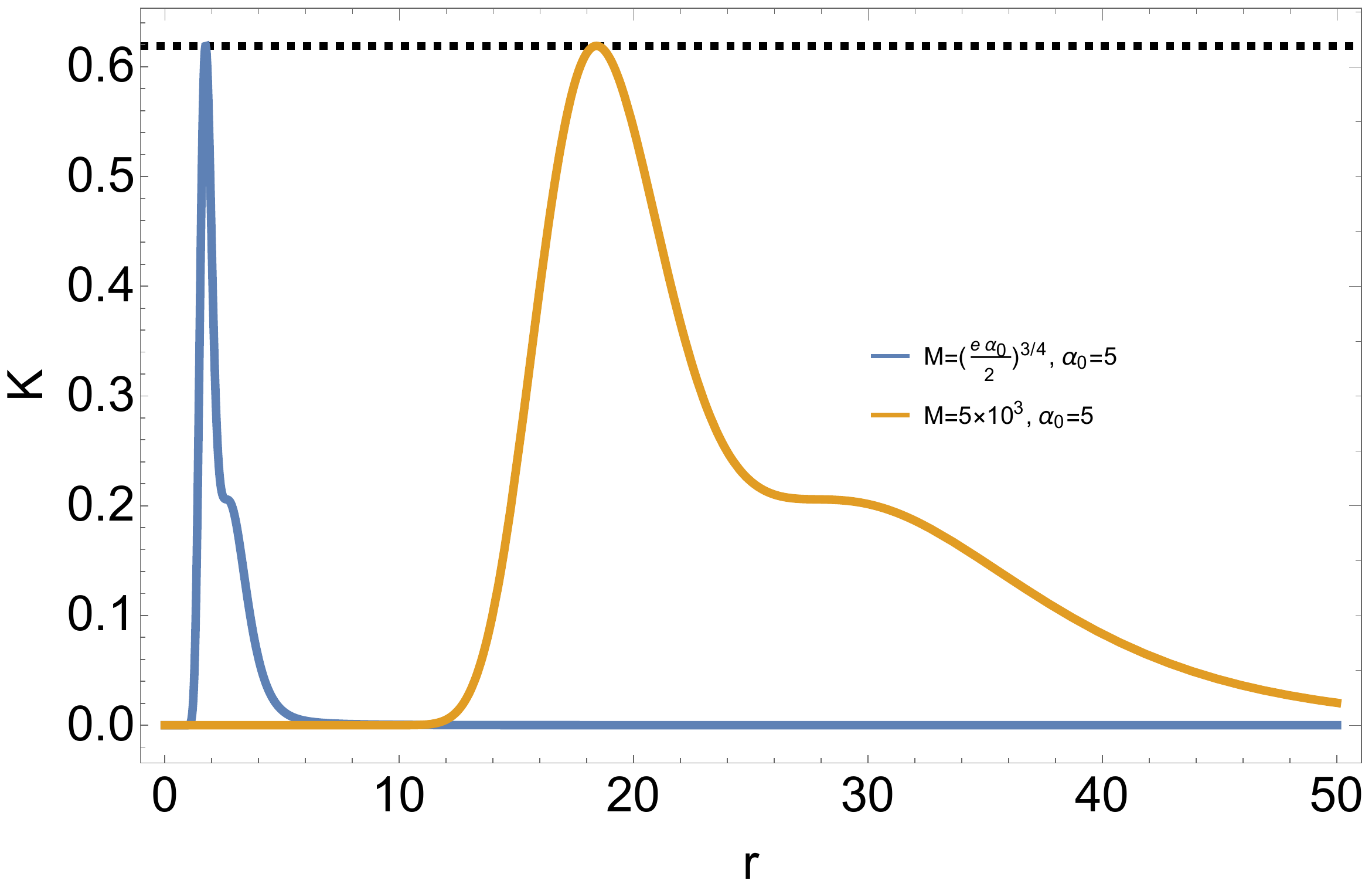}\ \hspace{0.05cm}
  \caption{\label{KB}  Kretschmann scalar curvature $K$  as the functions of the radial coordinate $r$ for Bardeen regular black hole(left) and $x=2/3$(right) respectively.}}
\end{figure}

\begin{figure} [t]
  \center{
  \includegraphics[scale=0.32]{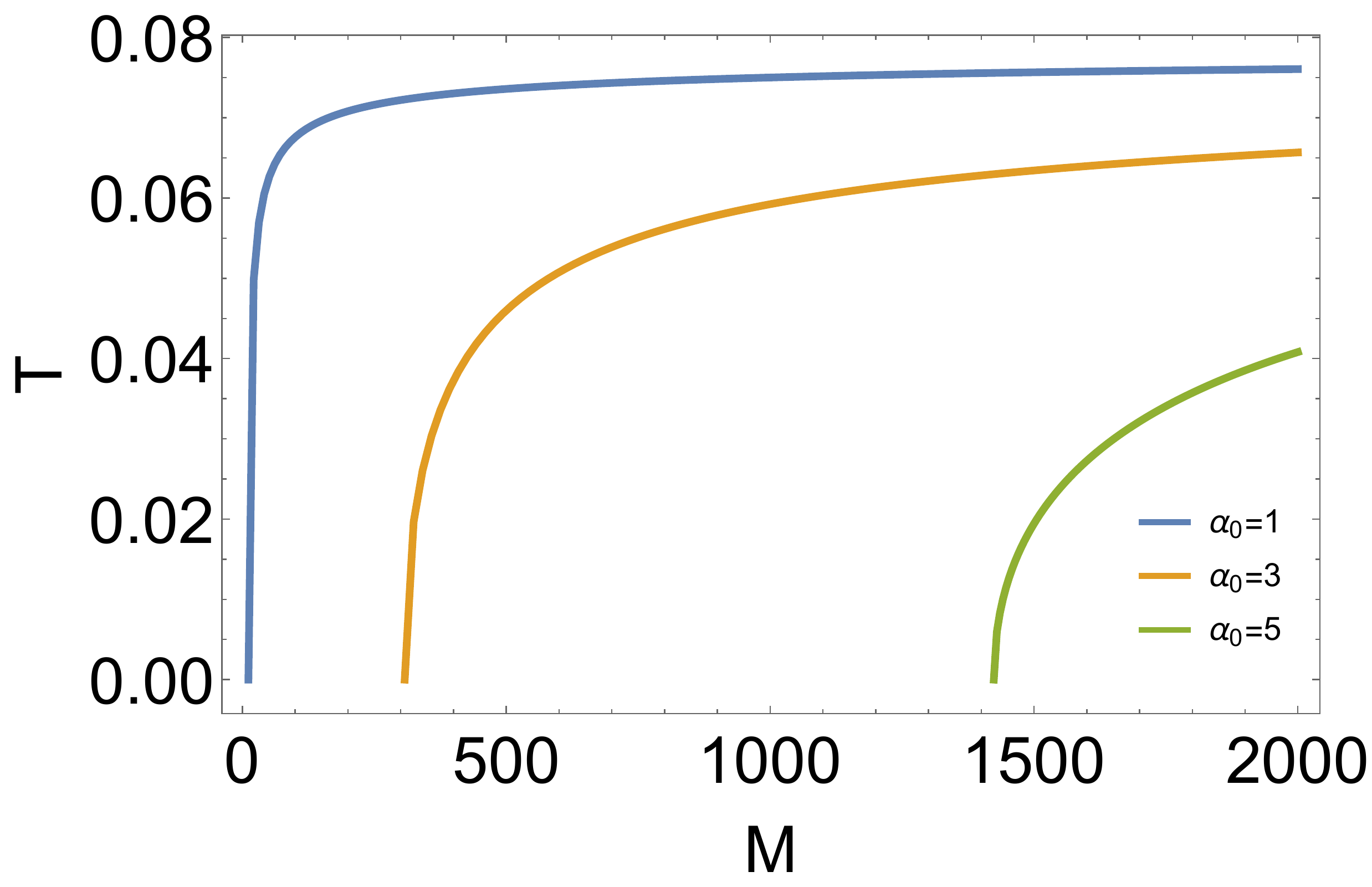}\ \hspace{0.05cm}
  \includegraphics[scale=0.425]{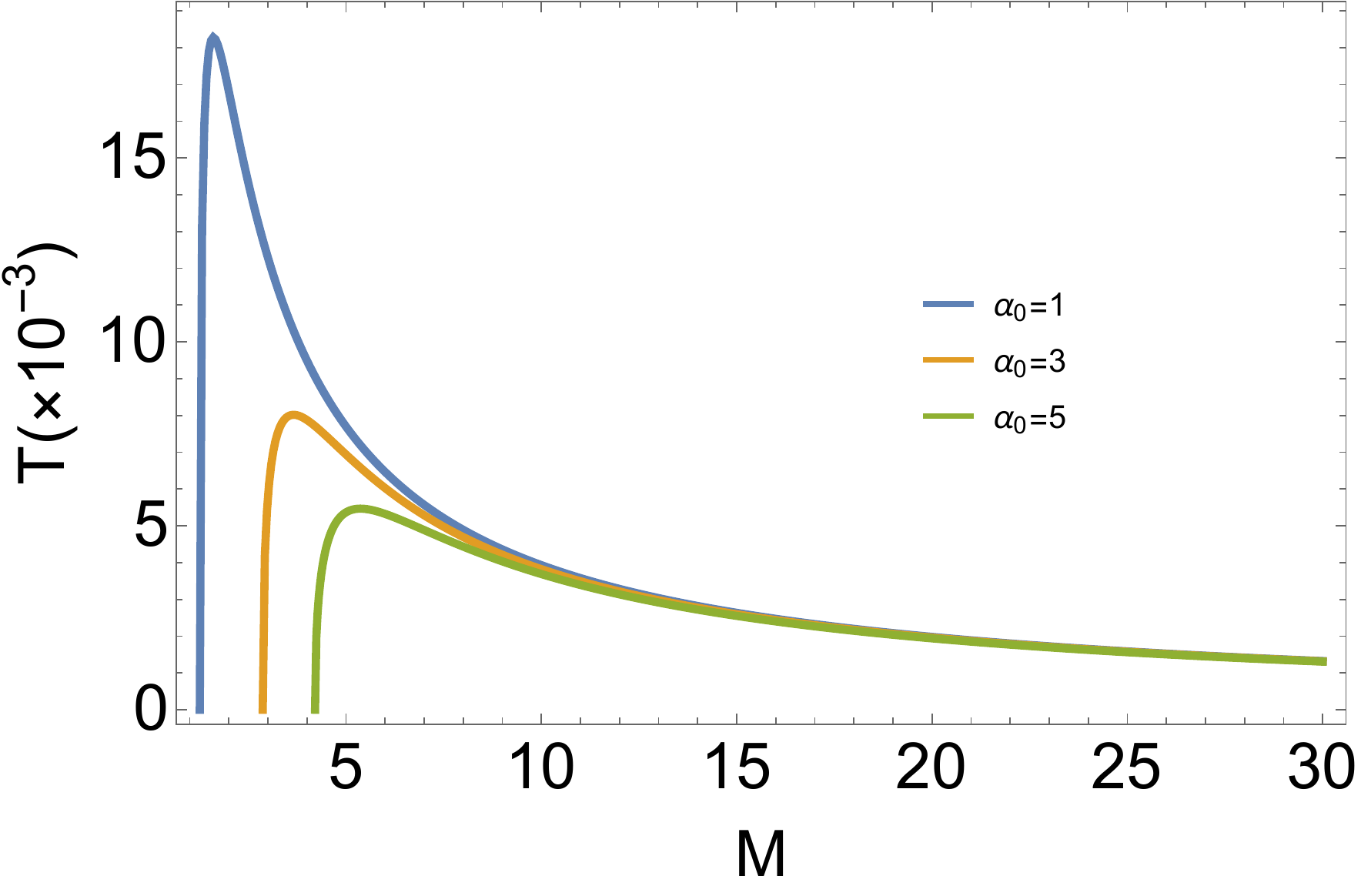}\ \hspace{0.05cm}
  \caption{\label{fig6} The Hawking temperature $T$  as the functions of $M$ for Bardeen black hole (left) and for the regular black hole with $x=2/3$(right).}}
\end{figure}

\begin{itemize}
\item The features of Kretschmann scalar curvature $K$. For
Bardeen black hole with the above potential, Kretschmann scalar
curvature is given by
\begin{equation}
\begin{aligned}
K &=\frac{1296 M^2 \left(32 \alpha _0^4 M^{8/3}-162 \alpha _0
M^{2/3} r^6+423 \alpha _0^2 M^{4/3} r^4-24 \alpha _0^3 M^2 r^2+81
r^8\right)}{\left(2 \alpha _0 M^{2/3}+3 r^2\right){}^7}.
\end{aligned}
\end{equation}
We plot Kretschmann scalar curvature $K$  as the function of the
radius for these regular black holes in Fig.\ref{KB}. Obviously,
for both black holes the maximum value of $K$ is independent of
mass $M$. The location of $K_{max}$ is always fixed at the center
for Bardeen black hole, independent of the parameter $\alpha_0$,
while for regular black hole with $x=2/3$, $K$ is always zero at
the center, and the location of $K_{max}$ moves to larger radius
with the increase of the parameter $\alpha_0$.

\item The features of thermodynamics. The Hawking temperature $T$
of Bardeen black hole is given by
\begin{equation}
T=\frac{3 \sqrt{3} M \left(3 r_h^3-4 \alpha _0 M^{2/3}
r_h\right)}{2 \pi  \left(3 r_h^2+2 \alpha _0
M^{2/3}\right){}^{5/2}}.
\end{equation}
Moreover, to form a black hole with horizon, the mass of the black
hole is bounded from below
\begin{equation}
M\geq \left(\frac{9 \alpha _0}{4}\right){}^3.
\end{equation}
When this bound is saturated, the black hole is characterized by
the minimal radius $r_h=\frac{3}{2} \sqrt{3 \alpha _0^3}$.

We plot the Hawking temperature $T$  as the functions of $M$ for
two black holes in Fig.\ref{fig6}, respectively. Both black holes
are characterized by a vanishing temperature at the final stage of
evaporation. However, for Bardeen black holes, we remark that
there is no maximal value for the Hawking temperature and  its
capacity exhibits a monotonic behavior with the mass with
$C=dM/dT>0$.

\end{itemize}

\section{The regular black hole corresponding to Hayward black hole at large scale}

In a quite parallel way, we construct a regular black hole
corresponding to Hayward black hole at large scales. With the
general form of gravity potential as Eq.(\ref{GFGP}), we find the
maximal value of Kretschmann scalar curvature $K$ behaves as
$K\propto \frac{M^2}{(\alpha_0M^x)^{6/n}}$. Thus if we set $n=3$,
then $x=1$ leads to a mass-independent curvature as well. In this
section we demonstrate that this regular black hole has the same
behavior as Hayward black hole at large scales indeed.

We start with the gravitational potential $\psi(r)$ as
\begin{equation}\label{Eq_grav}
\psi=-\frac{M}{r} e^{-\alpha / r^{3}}.
\end{equation}
Then Kretschmann scalar curvature is given by
\begin{equation}
\begin{aligned}
K &=\frac{12 e^{-\frac{2 \alpha }{r^3}} M^2}{r^{6}}(4
-\frac{32\alpha} {r^3}  +\frac{132 \alpha ^2}{r^6}-\frac{108
\alpha ^3}{r^9}+\frac{27 \alpha ^4}{r^{12}}).
\end{aligned}
\end{equation}
If we set $\alpha=\alpha_0 M $, then it goes to the curvature with
$x=1$. The maximal value of $K$ is proportional to $1/\alpha_0^2$.
The radius of the horizon $r_h$ is given by
\begin{equation}
r_h=2 M \left(\frac{\theta}{W_0(\theta)}\right)^{1/3}, \quad
\theta=-\frac{3\alpha_0 }{8 M^{2}}.
\end{equation}
The mass of the black hole is bounded as
\begin{equation}
M\geq \frac{1}{2} \sqrt{\frac{3 e \alpha _0}{2}},
\end{equation}
and thus the minimal radius of the horizon is
$r_h=\sqrt{\frac{3}{2} \alpha _0 e^{1/3}}$. The Hawking
temperature $T$ is given by
\begin{equation}
T=\frac{r_h^3-3 \alpha _0 M}{4 \pi  r_h^4}=\frac{1+W_0}{8 M \pi} \left( \frac{W_0}{\theta  }\right)^{1/3}.
\end{equation}

Similar to the analysis in previous section, we show that this
regular black hole reproduces the Hayward metric at large scales.
On one hand, for large $r\gg (\alpha_0M)^{1/3}$, the function
$F(r)$ in the metric behaves as
\begin{equation}
 F(r)=1+2 \psi(r)= 1-\frac{2M}{r} e^{-\alpha_0 M  / r^{3}}\cong 1-\frac{2M}{r} (1-\frac{\alpha_0 M}{r^{3}}+...).
\end{equation}
On the other hand, for Hayward regular black hole, the
gravitational potential $\psi(r)$ is specified as\footnote{We remark that when the dimension is restored, the expression of $\psi$ is $\psi=-\frac{GM r^2}{r^3+ \alpha _0 GM l_{p}^{2}}$. }
\begin{equation}
\psi=-\frac{M r^2}{r^3+ \alpha _0 M}.
\end{equation}
Therefore, at large scales the function $F(r)$ behaves
\begin{equation}
 F(r)=1+2 \psi(r)= 1-\frac{2M r^2}{r^3+M \alpha _0}\cong 1-\frac{2M}{r} (1-\frac{\alpha_0
M}{r^{3}}+...),
\end{equation}
which is identical to the regular black hole with $x=1$ and $n=3$
at large scale, indeed.

\begin{figure} [t]
  \center{
  \includegraphics[scale=0.31]{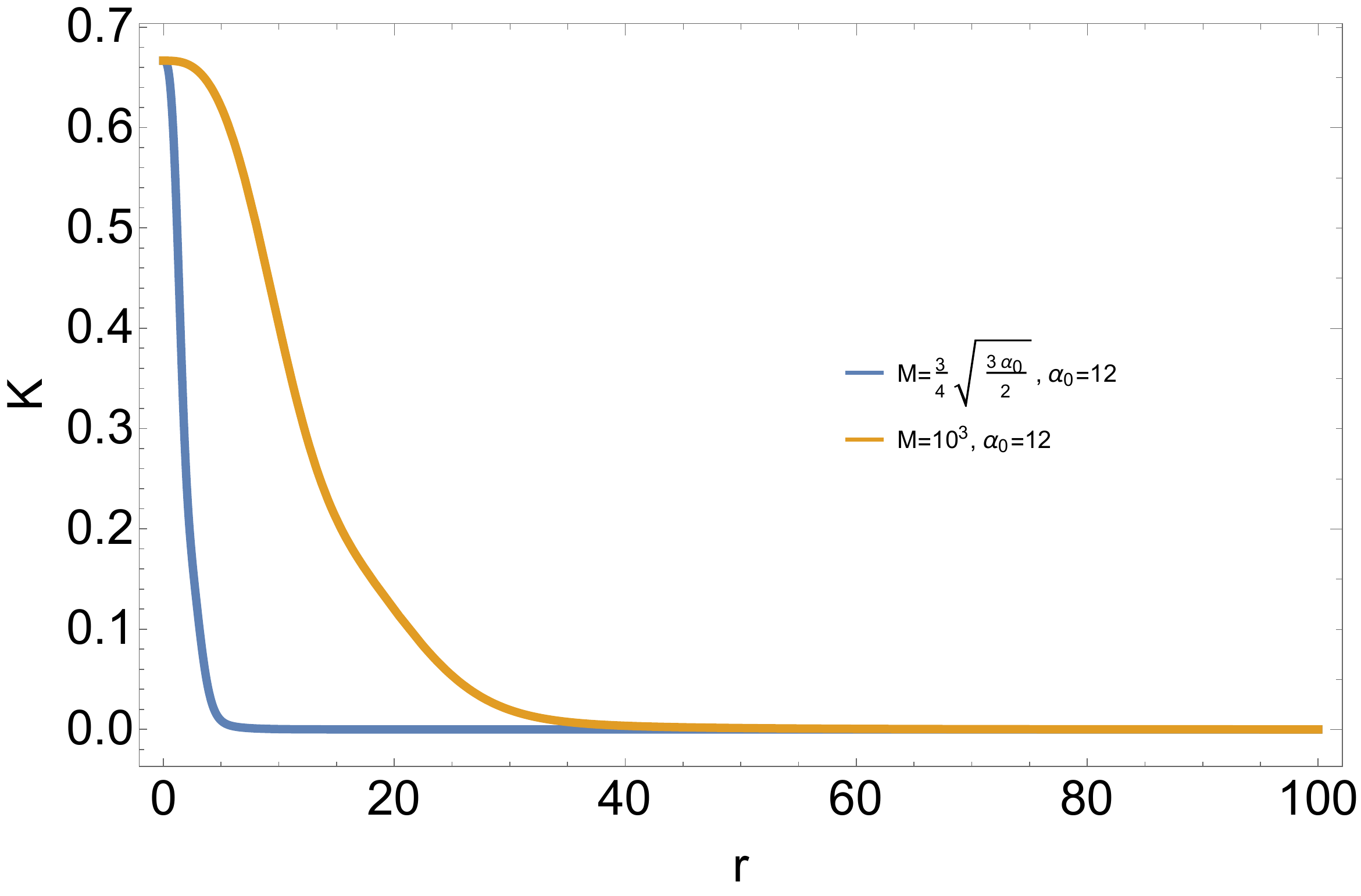}\ \hspace{0.05cm}
  \includegraphics[scale=0.32]{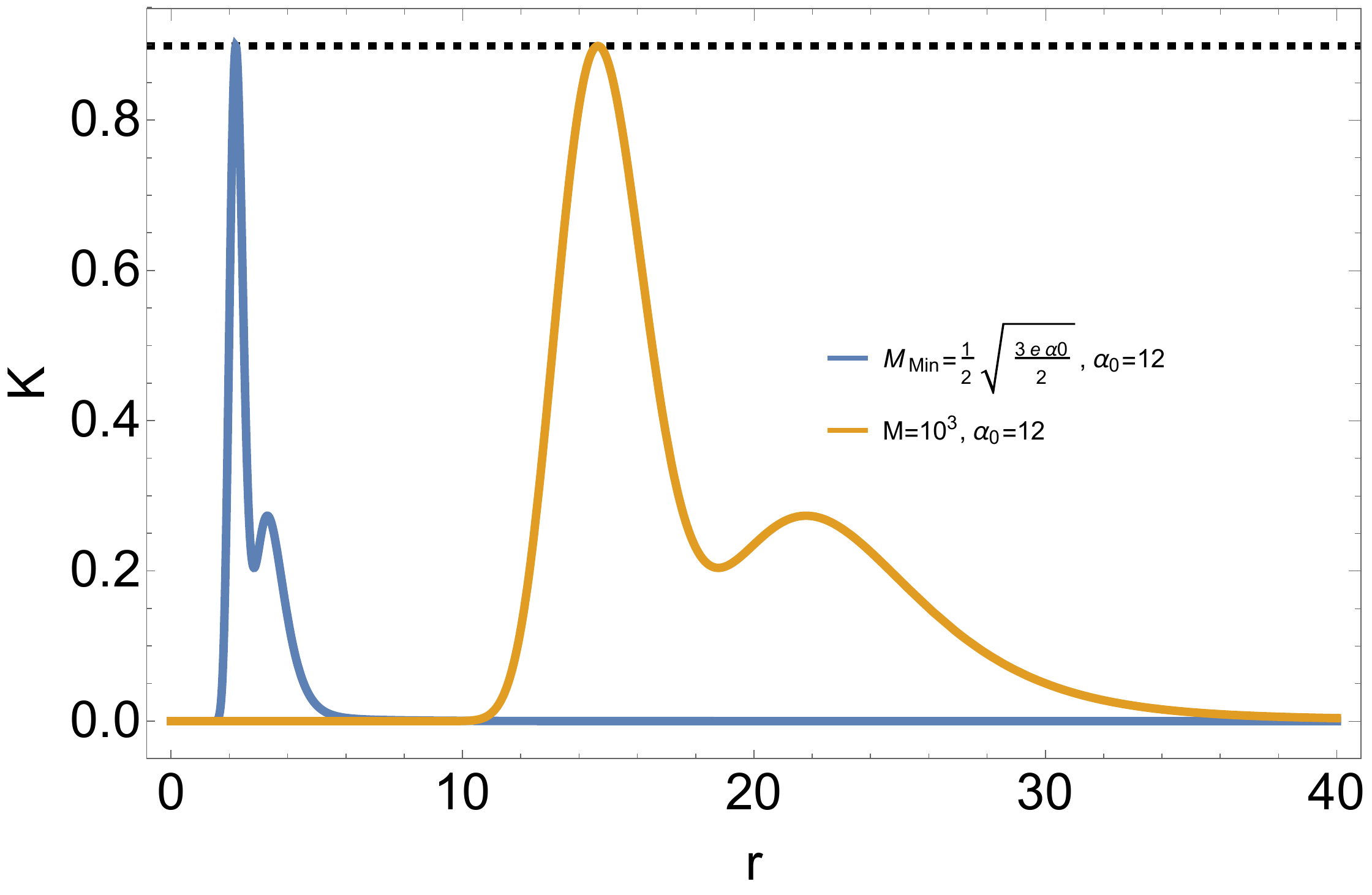}\ \hspace{0.05cm}
  \caption{\label{KBH}  The Kretschmann scalar curvature $K$  as the functions of the radial coordinate $r$ for Hayward regular black hole(left) and regular black hole described by Eq.(\ref{Eq_grav})(right) respectively.}}
\end{figure}

\begin{figure} [t]
  \center{
  \includegraphics[scale=0.4]{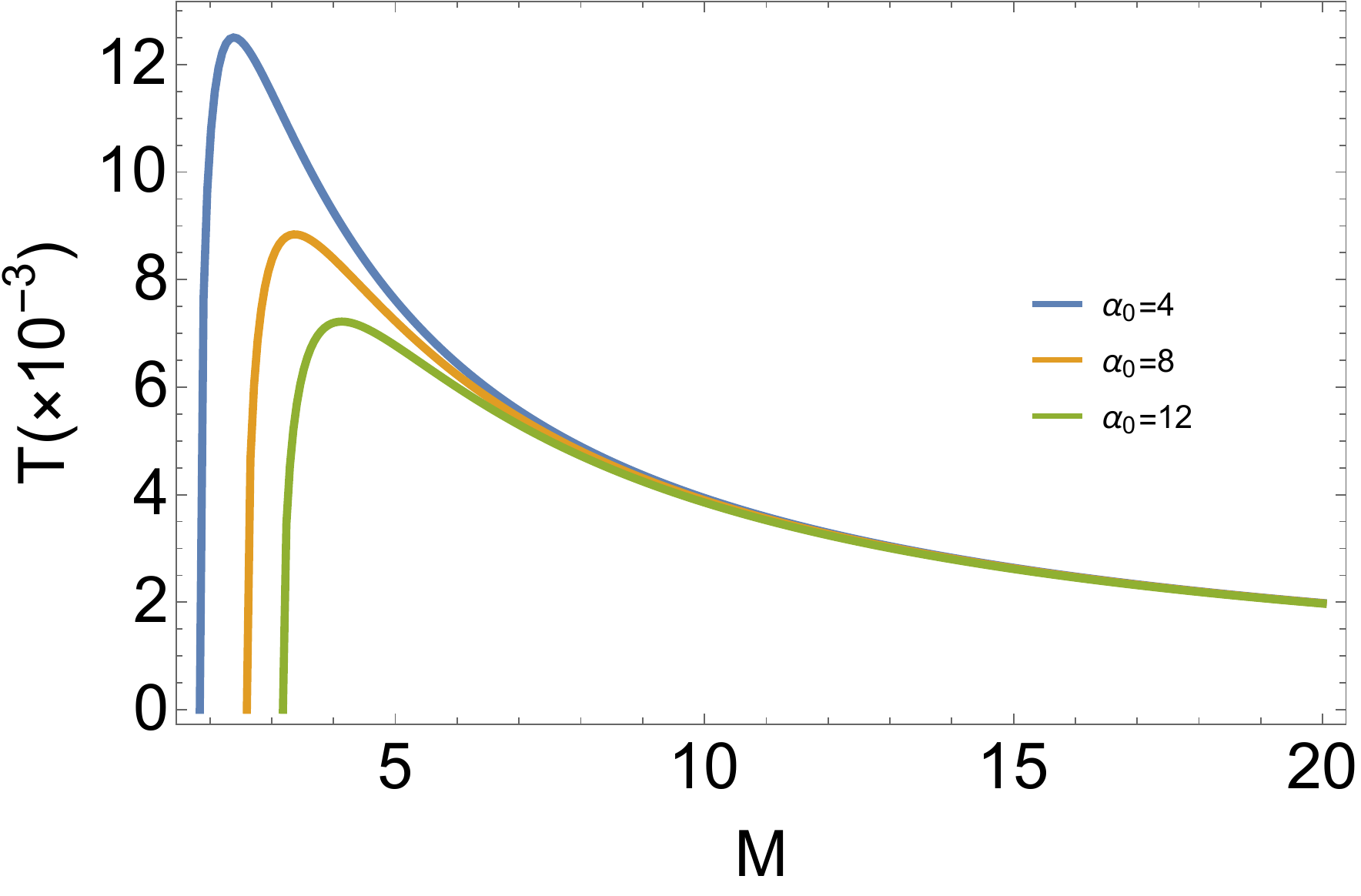}\ \hspace{0.05cm}
  \includegraphics[scale=0.4]{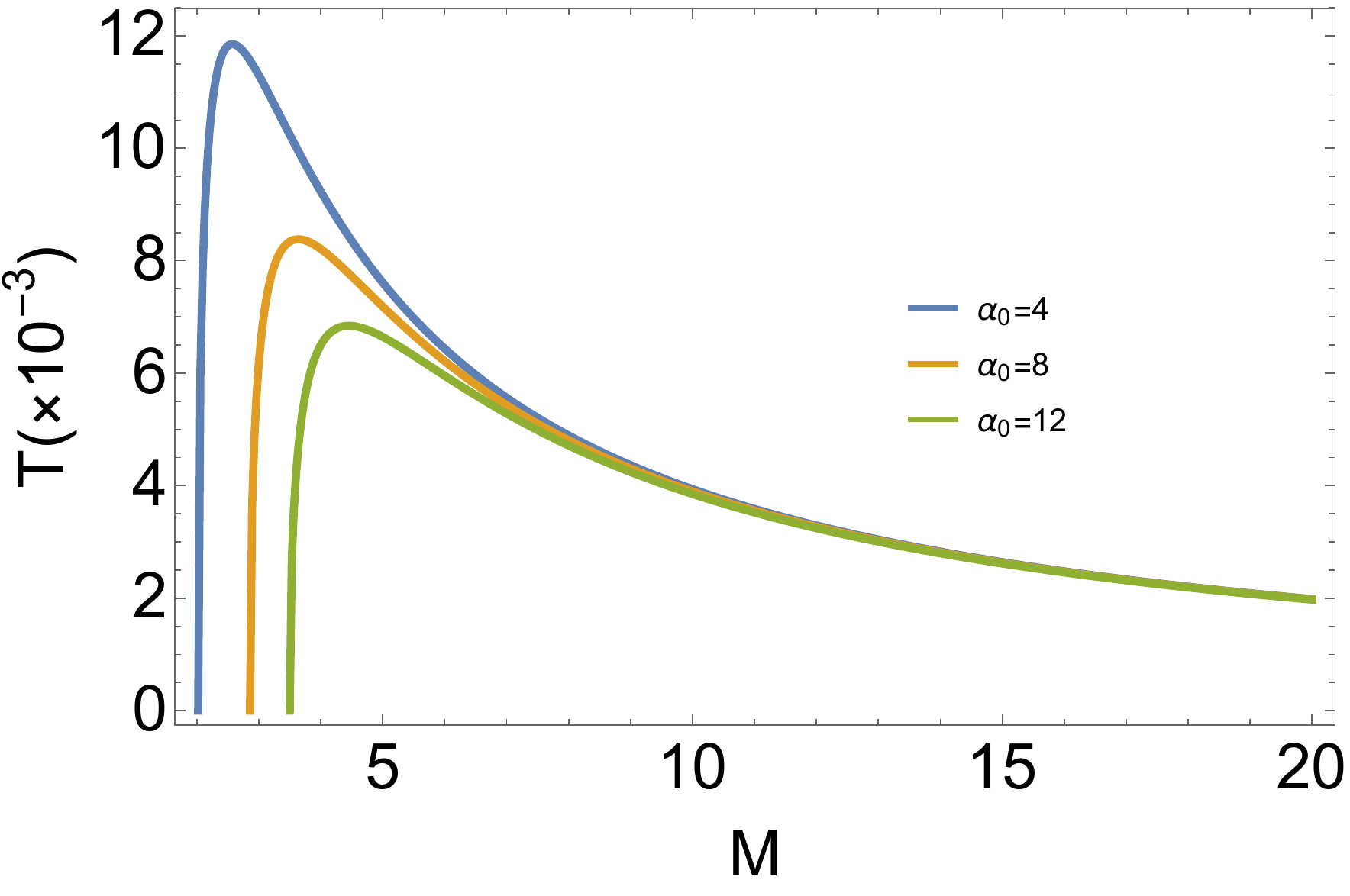}\ \hspace{0.05cm}
  \caption{\label{fig5}   The Hawking temperature $T$  as the functions of  $M$ for Hayward regular black hole(left) and regular black hole described by Eq.(\ref{Eq_grav})(right) respectively.}}
\end{figure}

As well, we compare the features of these two black holes as
follows.
\begin{itemize}
\item The features of Kretschmann scalar curvature $K$. For
Hayward black hole, Kretschmann scalar curvature is given by
\begin{equation}
\begin{aligned}
K &=\frac{48 M^2 \left(r^{12}-4 M r^9 \alpha _0+18 M^2 r^6 \alpha
_0^2-2 m^3 r^3 \alpha _0^3+2 M^4 \alpha _0^4\right)}{\left(r^3+M
\alpha _0\right){}^6}.
\end{aligned}
\end{equation}
We plot Kretschmann scalar curvature $K$  as the function of the
radius for these regular black holes in Fig.\ref{KBH}. Again, for
both black holes the maximum value of $K$ is independent of mass
$M$. The location of $K_{max}$ is always fixed at the center for
Hayward black hole, independent of the parameter $\alpha_0$, while
for regular black hole with $x=1$, $K$ is always zero at the
center, and the location of $K_{max}$ moves to larger radius with
the increase of the parameter $\alpha_0$.

\item The features of thermodynamics.  The mass of Hayward black
hole is bounded by
\begin{equation}
M\geq \frac{3}{4} \sqrt{\frac{3 \alpha _0}{2}}.
\end{equation}
The minimal radius of the horizon is $r_h=\sqrt{\frac{3 \alpha
_0}{2}}$. The Hawking temperature $T$ is given by
\begin{equation}
T=\frac{M r_h \left(r_h^3-2 \text{M$\alpha $}_0\right)}{2 \pi
\left(r_h^3+M \alpha _0\right){}^2}.
\end{equation}

We plot the Hawking temperature $T$ as the functions of $M$ for
two black holes in Fig.\ref{fig5}, respectively. Both black holes
are characterized by a vanishing temperature at the final stage of
evaporation.  It is also interesting to notice that the mass
dependent behavior of the temperature is quite similar for these
two black holes, even at the final stage of the evaporation.

\end{itemize}

\section{The regular black hole with general exponential form }
In this section, we consider the generalized form of the black
hole with the gravitational potential as specified in Eq.
(\ref{GFGP}). In Tab.\ref{Tabel1}, we summarize the dependent
behavior of different quantities on $\alpha _0$ with factors $x$
and $n$. We emphasize that $\alpha _0$ is dimensionless and does not depend on $x$ for given $n$.

\begin{table*}[t]
\newcommand{\tabincell}[2]{\begin{tabular}{@{}#1@{}}#2\end{tabular}}

\begin{ruledtabular}
\begin{tabular}{ccccc}
$K_{Max}$ & $r_{h \ Min }$  & $M_{Min}$ & $T_{Max}$ & $M_{T_{Max}}$ \\  \hline
$M^{2-6x/n}\alpha _{0}^{-6/n}$ & $\alpha _{0}^{1/(n-x)}$  & $\alpha _{0}^{1/(n-x)}$ & $\alpha _{0}^{1/(x-n)}$ & $\alpha _{0}^{1/(n-x)}$ \\
\end{tabular}
\end{ruledtabular}
\caption{\label{Tabel1} The dependent behavior of different
quantities on $\alpha _0$, where $r_{h \ Min}$  and $M_{Min}$ are the
minimal values of the horizon and the mass at $T=0$ respectively,
while $M_{T_{Max}}$ is the mass corresponding to $T=T_{Max}$.}
\end{table*}

The special case happens whenever $x=n$. It is obvious to see that
the mass of the black hole is not bounded by $\alpha_0$ any more,
and the temperature becomes divergent when  mass goes to zero,
just like the Schwarzschild black hole. Thus, to form a regular
black hole with remnant, we require that $x<n$. Moreover, if we
expect the maximal value of Kretschmann scalar curvature $K$ is
bounded above with the increase of the mass $M$, we find $x\geq
n/3$. In particular, when $x=n/3$, $K_{max}$ is mass independent.
Thus, we conclude that to form a regular black hole with
sub-Planckian Kretschmann scalar curvature, we demand $n>x\geq
n/3$.

On the other hand, given the regular black hole with the
gravitational potential as specified in Eq.(\ref{GFGP}), we may
construct the corresponding regular black hole with asymptotically
de-Sitter core which has the same behavior at large scales. The function
$F(r)$ in the metric reads as
\begin{equation}
 F(r)=1+2 \psi(r)=1-\frac{2 M r^{\frac{n}{x}-1}}{\left(r^n+x \alpha_0  M^x\right)^{1/x}},
\end{equation}
which can be viewed as the generalization  of Bardeen and Haywald
black holes. In addition, we may give a remark about Frolov black
hole and its correspondence. For Frolov black hole, the
gravitational potential reads as
\begin{equation}
 \psi(r)=1-\frac{2 M r^2}{\alpha_0 ^3+\alpha_0  M+r^3}.
\end{equation}
 At large scale, the black hole is identical to the regular black
hole with potential $\psi(r)=
-\frac{M}{r}e^{-\frac{(\alpha_0M+\alpha_0^3)}{r^3}}$.

Finally, we briefly discuss the effective stress-energy tensor and
the energy condition. The effective stress-energy tensor
corresponding to Eq.(\ref{GFGP}) is given by
\begin{equation}
\begin{aligned}
T_{0}^{0}&=T_{1}^{1}=\frac{1}{4\pi r^2}\alpha _0 n M^x r^{-n} \psi (r), \\
T_{2}^{2}&=T_{3}^{3}= \frac{\alpha _0 n M^x r^{-2 (n+1)} \psi
   (r) \left(\alpha _0 n M^x+(-n-1)
   r^n\right)}{8 \pi }.
\end{aligned}
\end{equation}
The strong energy condition (SEC) is violated when
$-T_{0}^{0}+T_{1}^{1}+T_{2}^{2}+T_{3}^{3}<0$, namely
$r<\left(\frac{ \alpha _0 n M^x}{n+1}\right)^{\frac{1}{n}}$. We
notice that only for $x=0$, the region with the violation of SEC
is mass independent, while for other allowed values of $x$, the
region becomes larger with the increase of the mass.

\section{Heuristical understanding on modified gravity potential}

In this section we will present a heuristical understanding on the
origin of such exponentially suppressing potential, closely
following the strategy proposed in
\cite{Xiang:2013sza,Li:2016yfd}. As pointed out in
\cite{Li:2016yfd}, when the quantum effects of gravity are taken
into account, the same observed results such as COW phase shift in
gedanken experiment\cite{Colella1975} can be understood by two equivalent
pictures. One is that the classical
gravitational field strength remains unchanged by quantum object,
but the usual Heisenberg's uncertainty principle obeyed by the
object is forced to be a generalized one
\begin{equation}
[\hat{x},\hat{p}]=iz(\hat{p}),
\end{equation}
where $z(\hat{p})$ could be a general function of the momentum
operator which reflects the effects of gravity on this quantum
object. Typically, the widely considered one in literature is
$z(\hat{p})=1+\alpha \hat{p}^2 $, which leads to a quantum theory
with the minimal observable length.  {\it Alternatively}, one can
assume that the quantum objects still obey the usual Heisenberg's
uncertainty relation, namely the quantum theory is retained, but
introduce an effective gravitational field strength to count in
the interacting effects of gravity and the quantum object. It
turns out that in latter point of view,  the space time with
Schwarzschild metric will be modified to be one characterized by
an effective Newton constant $G'=G/z$. Namely, we obtain a
modified Schwarzschild metric as \cite{Xiang:2013sza,Li:2016yfd}
\begin{equation}\label{ENC}
d s^{2}=- (1-\frac{2MG}{r^2 z}) d t^{2}+(1-\frac{2MG}{r^2 z})^{-1} d r^{2}+r^{2} d
\Omega^{2},
\end{equation}
which becomes the starting point that we propose the modified
metric for regular black holes in this paper.  However, rather
than considering the ordinary GUP associated with
$z(\hat{p})=1+\alpha \hat{p}^2 $,  we introduce a general
exponential function with $z(\hat{p})=e^{\alpha \hat{p}^a}$, which
has also previously considered in \cite{Harbach:2005yu}.
Obviously, when $a=2$ and $\alpha p^2\ll 1$, this goes back to
$z(\hat{p})=1+\alpha \hat{p}^2+... $. In this sense, such an
exponential form includes the non-perturbative effects of quantum
gravity because in the expansion of weak momentum it recovers the
quadratic form of the momentum.

Next, we need to evaluate the momentum uncertainty of any quantum
object with $m$ in the gravitational field. We know,  for an
observable quantity, $p^2\geq\triangle p^2$. The key point is that
whenever the quantum effects of gravity are taken into account, the
quantum probe must be characterized by a position uncertainty
bounded below by the Planck length, namely  $\triangle x\geq1$.
Therefore, it must experience the gravitational tidal force in
curved spacetime, which leads to

\begin{equation}
(\triangle p)^2\geq \frac{\triangle p}{\triangle x}=\frac{F\triangle t}{\triangle x}=\frac{2GM m\triangle t\triangle
x}{\triangle x r^3}\geq \frac{ 2GM\triangle E\triangle t}{r^3}\geq \frac{ 2GM}{r^3},
\end{equation}
where $\triangle E$ and $\triangle t$ are the characteristic
energy of the probe like a photon and the time in the process of
detection, which is supposed to have a photon-particle collision.
To avoid new particle pair production, $\triangle E\leq m$ is assumed.
Thus, we have
\begin{equation}
z=e^{\alpha p^a}\sim e^{\alpha (\frac{ 2GM}{r^3})^{a/2}}.
\end{equation}
In particular, if we set $\alpha=\alpha_0/2$ and $a=2$, then
by identifying $G'=G/z$ and plugging it into the metric, we
find the corresponding metric is nothing but the one of regular
black hole with $n=3$ and $x=1$. Or, if we set
$\alpha=2^{3/2}\alpha_0$ and $a=4/3$, then it exactly gives rise
to the regular black hole with $n=2$ and $x=2/3$. In general if
we set $\alpha\propto \alpha_0 M^{x-n/3}$ and $a=2n/3$, then
\begin{equation}
z=e^{\alpha p^{2n/3}}\sim e^{\frac{\alpha_0 M^{x}}{r^n}},
\end{equation}
leading to the potential given in Eq.(\ref{GFGP}).
In parallel, if one considers the leading term as the
ordinary GUP, then
\begin{equation}
z=1+\alpha p^{2n/3}\sim 1+\frac{\alpha_0 M^{x}}{r^n}.
\end{equation}
Identifying $G'=G/z$ and plugging it into the metric, one obtains
a sort of regular black holes with de-Sitter core including
Bardeen black hole as well as Hayward black hole.

\section{Conclusion and Discussion}
In this paper we have introduced an exponentially suppressing
gravitational potential to construct a sort of regular black holes
with asymptotically Minkowski core. In contrast to all previous
regular solutions with Minkowski core, the Kretschmann scalar
curvature is not only finite everywhere, but also bounded above by
the Planck mass density, regardless of the mass of the black hole.
Without doubt, from the viewpoint of quantum gravity, the
spacetime with sub-Planckian curvature is more realistic and
applicable to the whole process of evaporation. We think this is a
dramatic improvement in comparison with the previous regular black
holes with Minkowski core.
In the thermodynamical aspect, all
these regular black holes are characterized by a vanishing Hawking
temperature at the late stage of evaporation and the remnant has a
minimal mass at the Planck scale, which may be viewed as the
candidate of dark matter.
As shown in \cite{Adler:2001vs,Dymnikova:2007gx,Dymnikova:2015yma}, the remnant of regular black holes may be viewed as a candidate for dark matter. The contribution of the remnant characterised by $\alpha_0$ is constrained by the observed density of dark matter. Thus, the more detailed detection of dark matter would potentially provide constraints on the choices of the parameter value $\alpha_0$ in future.
Furthermore, we have demonstrated that for specifical choice of
the potential from, the regular black holes have the similar
behavior as Bardeen/Hayward/Frolov black holes at large scales,
respectively. Therefore, we have established a one-to-one
correspondence between the regular black holes with Minkowski core
and those with de-Sitter core. This correspondence provides us a
scheme to construct new regular black holes with de-Sitter core as well.  Theoretically, these two different sorts of black holes may be ascribed to the different forms of the modified gravitational potential, which are supposed to contain some non-perturbative corrections due to quantum gravity effects such that the singularity could be avoided. We remark that these effects should be non-perturbative in the sense that the modified gravitational potential could be expanded as the polynomials of the distance and the leading term is the standard classical gravitational potential, similar to 1-loop corrections in perturbative calculations\cite{Bjerrum-Bohr:2002gqz,Donoghue:2012zc,DeLorenzo:2014pta,LingLingYi:2021rfn}. But 1-loop correction is obtained with perturbative method and is not strong enough to get rid of the singularity. Moreover, the exponential form for black holes with Minkowski core implies that regular black holes with Minkowski cores have stronger non-perturbative corrections compared to those with de-Sitter cores.
 We comment that in this paper we have mainly considered the parameters leading to the Bardeen/Hayward/Frolov black hole at large scales such that  the Kretschmann scalar
curvature of the regular black hole with a Minkowski core has the same feature as those with de-Sitter core, namely $K_{max}$ is mass independent. Definitely, one may consider the regular black holes with other parameters whose Kretschmann scalar
curvature may be mass dependent but maintain sub-Planckian.
We also remark that the choices of these parameters are not random, but different choices do imply there are different forms of the modified gravitational potential due  to the quantum gravity effects.  The precise form of the correction terms remains to be studied through further research on quantum theory of gravity and its phenomenology.
In addition, the regular black hole with de-Sitter core has a Kretschmann scalar curvature that reaches its maximum value at the center of the black hole, while the regular black hole with Minkowski core does not. Instead, the Kretschmann scalar curvature is always zero at the center and reaches its maximum value at a point between the center and the horizon. This distinction may result in observable effects in the future  and has been theoretically investigated in \cite{Zeng:2022tzh,Ling:2022vrv}.  It is found that the shadows of black holes with Minkowski cores have larger deformations than those with de-Sitter cores. Moreover, the radius of the photon sphere in regular black hole with  de Sitter cores is larger than the black hole with  Minkowski cores.  It is desirable to  distinguish these  two types of black holes by observation in future.

Finally, we have provided a theoretical understanding on the
origin of such exponentially suppressing potential appeared in the
black hole metric in a heuristic manner. Two essential ingredients
are taken into account. Firstly, any quantum object has a size
larger than the minimal length at Planck scale due to the effects
of gravity. Secondly, any object with a finite size must be
affected by the gravitational tidal force. Moreover,  the
exponentially suppressing form of the gravity potential in the
metric results from the generalized function $z(p)$ with a novel
exponential form of the momentum in GUP, which is in contrast to
the ordinary one with the quadratic form of the momentum. This
generalization may reflect the non-pertubative effects of quantum
gravity. It is also this relation that paves a bridge between the
effective metric of regular black holes with asymptotically
Minkowski core with that with asymptotically de-Sitter core.

Of course, in this paper the regular black holes that we have
constructed are static, we expect the Hawking radiation may be
added in such that they can be extended to dynamical ones such as
Vaidya-like solutions to investigate the formation and evaporation
of regular black holes with Minkowski core, as demonstrated for
the black hole with de-Sitter core in \cite{Hayward:2005gi}. The whole
picture about the black hole evaporation will help us to
understand the final fate of black hole evolution and disclose the
mystery of black hole information loss paradox.

\section*{Acknowledgments}

We are very grateful to Prof. Xiang Li for helpful discussions.
This work is supported in part by the Natural Science Foundation
of China under Grant No.~11875053 and 12035016.  It is also supported by Beijing Natural Science Foundation under Grant No.~1222031.

\end{document}